\documentclass[11pt,letterpaper]{article}
\usepackage{amsmath, amsfonts}
\usepackage{epsfig}
\usepackage{array}
 \newcommand{\psid}{\ensuremath{\psi^\dagger}}
\newcommand{\er}{\ensuremath{\hat{e}_r}}
\newcommand{\eth}{\ensuremath{\hat{e}_\theta}}
\newcommand{\eph}{\ensuremath{\hat{e}_\phi}}
\newcommand{\half}{\ensuremath{\frac{1}{2}}}
\newcommand{\va}{\ensuremath{{\bf v}_a}}
\newcommand{\vb}{\ensuremath{{\bf v}_b}}
\newcommand{\partl}[2]{\ensuremath{\frac{\partial{#1}}{\partial{#2}}}}
\newcommand{\sch}{Schr\"{o}dinger }
\newcommand{\be}{\begin{equation}}
\newcommand{\ee}{\end{equation}}
\begin{document}
\title{Spin-dependent Bohm trajectories for Pauli and Dirac
eigenstates of hydrogen}
\author{C. Colijn and E.R. Vrscay
\\Department of Applied Mathematics\\University of Waterloo
\\Waterloo, Ontario, Canada  N2L 3G1}
\maketitle

\begin{abstract}
The de Broglie-Bohm causal theory of quantum mechanics is applied to
the hydrogen atom in the fully spin-dependent and relativistic
framework of the Dirac equation, and in the nonrelativistic but
spin-dependent framework of the Pauli equation.  Eigenstates are
chosen which are simultaneous eigenstates of the energy $H$, total angular
momentum $M$, and $z$ component of the total angular momentum
$M_z$. We find the trajectories of the electron, and show that in these
eigenstates, motion is circular about the $z$-axis, with constant angular velocity.
We compute the rates of revolution for the ground
($n=1$) state and the $n=2$ states, and show that there is agreement
in the relevant cases between the Dirac and Pauli results, and with
earlier results on the Schr\"{o}dinger equation.
\end{abstract}
\begin{center}
\end{center}

\section{INTRODUCTION}
In Bohm's original causal interpretation of quantum mechanics \cite{Bo52},
the motion of a quantum particle is determined by its
\sch wave function $\psi$, which acts as a kind of guidance
wave \cite{deBr60}.  If the wave function is written as
\be
\psi({\bf x},t) = R({\bf x},t)e^{iS({\bf x},t)/\hbar},
\ee
where $R$ and $S$ are real-valued, then the trajectory of the particle
is determined by the guidance relation
\be
{\bf p}=\nabla S.
\label{basicbohm}
\ee
The momentum is related to the well-known \sch current {\bf j} as
follows,
\be
\label{currvel}
{\bf p} = \frac{m}{\rho}{\bf j}  \end{equation}
where $\rho = \psid\psi = R^2$.
Comprehensive discussions of the de Broglie-Bohm causal interpretation
can be found in \cite{BoHi} and \cite{Ho93}.

It is quite natural to examine the hydrogen atom, one of the simplest
quantum systems, in terms of the de Broglie-Bohm theory.
Indeed, as originally discussed in \cite{Bo52}, the \sch
guidance relation \eqref{basicbohm} predicts that ${\bf p} = {\bf 0}$
for all real eigenstates, including the ground and all higher
$s$ hydrogenic states.
However, as Holland \cite{Ho99} pointed out, Eq. \eqref{basicbohm} is
valid only for spinless particles.  For particles with spin,
the condition of Lorentz covariance on the law of motion implies that
the momentum of a particle with spin {\bf s} must be given by
\be
\label{withterm}
{\bf p}=\nabla S + \nabla \log\rho \times {\bf s},
\ee
where $\rho=\psid\psi$ \cite{He79,Ho99}.
Only in this way can the theory be
embedded in a relativistic formulation.
Indeed, in papers \cite{GuHe,He75,He79},
it was shown that in order for it to be consistent
with Dirac theory, the \sch equation must be regarded
as describing an electron in a definite eigenstate of spin.
In these papers, the current vector associated with \eqref{withterm},
\be
\label{paulicurrent}
{\bf j} = \frac{1}{m}\rho \nabla S + \frac{1}{m} \nabla \rho \times {\bf s},
\ee
was referred to as the {\em Pauli current}, the nonrelativistic
limit of the {\em Dirac current}.

The guidance law \eqref{withterm} no longer implies that ${\bf p}={\bf 0}$
for real eigenstates so it is natural to ask how it applies
to the hydrogen atom.
In \cite{CoVr}, we showed that for an electron in a spin eigenstate with
$s_z = \pm \frac{1}{2}$, the spin-dependent term in \eqref{withterm}
will be responsible for a motion in a plane perpendicular to
the $z$-axis and along a contour of constant $\rho$ value.
For the case of an electron in a
Schr\"{o}dinger energy/angular momentum eigenstate, $\psi_{nlm}$,
this implies circular motion about the $z$-axis.

In this paper, we examine de Broglie-Bohm trajectories for
an electron in a hydrogen atom as described by the Pauli
and Dirac equations using appropriate (spin-dependent) currents.
The electron is assumed to be in a Pauli or Dirac eigenstate of energy
and total angular momentum.  Note that, in general, this does
not imply that the electron is in a spin eigenstate of known
$s_z$ value.
We also show that
under appropriate nonrelativistic limits, the angular rotation rates 
for Dirac trajectories become those of Pauli trajectories.
In the cases that the electron is in a spin eigenstate
(e.g., $1s$, $2s$, $2p_0$), the Pauli rotation rates agree
with the \sch trajectories obtained in \cite{CoVr}.
The result is a coherent 
application of de Broglie-Bohm theory
to relativistic and nonrelativistic hydrogen atom eigenstates.  

For both the Dirac and Pauli cases, the \sch guidance formula in
\eqref{withterm} can be generalized using the relationship
\eqref{currvel} where {\bf j} is the appropriate (Dirac/Pauli) current
and $\rho = \psid\psi$.
First, consider the Dirac equation,
\begin{equation}
 i\hbar \partl{}{t}\psi =
        ( -e \phi + \beta E_o + {\boldsymbol
	\alpha}\cdot(c{\bf p} + e{\bf A}) ) \psi  .
\end{equation}
Here, $\psi = (\psi_1,\psi_2,\psi_3,\psi_4)$ is a four-component
wave function,
$\phi$ and ${\bf A}$ are the scalar and vector potentials, $E_0$
and ${\bf p}$ are the rest mass energy and momentum operators, $e$ is
the electric charge, and $\boldsymbol\alpha$ and $\beta$ are the
Dirac operators.
In this study, ${\bf A} = {\bf 0}$ and the
current is given by
\begin{equation}\label{dcurrent} {\bf j}= c\psid {\boldsymbol
\alpha}\psi = (j_x, j_y, j_z),\end{equation} where the
${\boldsymbol \alpha}$ are the $4\times4$ Dirac matrices,
\be
{\boldsymbol  \alpha} = \begin{pmatrix} 0 &{\boldsymbol \sigma}\\
{\boldsymbol \sigma}&0\end{pmatrix}, 
\ee
and the ${\boldsymbol \sigma}$
are the $2 \times 2$ Pauli matrices.

If the particle is in a potential such that $e\phi \ll mc^2$, then
there exist stationary states for which the average velocity $\bar{v}$ is
nonrelativistic, and $E \approx E_0=mc^2$.
In this case, the latter two components of the 4-component Dirac state
are smaller in magnitude than the first
two components by a factor of $\bar{v}/c$.
The Dirac equation may then be reduced to the Pauli equation involving
the two components $\psi_1$ and $\psi_2$:
\begin{equation}\label{pauli}
i\hbar \frac{\partial\psi}{\partial t}=
\frac{1}{2m}(-i\hbar\nabla+e{\bf A})^2\psi +
\frac{e\hbar}{2m}{\boldsymbol \sigma}\cdot{\bf B} \psi -eV\psi.
\end{equation}
Here $\psi = (\psi_1,\psi_2)$ is a two-component Pauli spinor wave function.
Once again, we assume that ${\bf A} = {\bf 0}$
so that the associated Pauli current is given by \cite{BoHi}
\begin{equation}\label{pcurrent}
{\bf j}={\bf j}_A + {\bf j}_B =\frac{\hbar}{2mi}(\psi^\dagger\nabla\psi -
\psi\nabla\psi^\dagger)+\frac{\hbar}{2m}\nabla \times
(\psi^\dagger {\mathbf \sigma}\psi).
\end{equation}
Note that if one
assumes that the system is in an eigenstate of the spin operator, then
\eqref{pcurrent} and \eqref{currvel} together reduce to
\eqref{withterm}.

In the case of the \sch equation for hydrogen, it is usually assumed that spin
interactions are negligible so that the
wave function can be written as a product of spatial- and spin-dependent
terms, i.e.,
\be
\psi = \psi({\bf r}, t)\chi_s . 
\ee
As is well known, the spatial hydrogenic energy \sch eigenfunctions,
\be
 \psi_{n,\ell, m}(r,\theta,\phi)= R_{n,\ell}(r)Y_{lm}(\theta,\phi),
\ee
solutions to the (spinless) time-independent \sch equation, are
also simultaneous eigenstates of the orbital angular
momentum operator $L^2$, with eigenvalues $\hbar^2\ell(\ell+1)$, and the
operator $L_z$, with eigenvalues
$\hbar m$.

In the Pauli equation, where spin-orbit interactions are
excluded, the orbital angular momentum operator $L^2$
commutes with the hamiltonian. This is not the case for
the Dirac equation.  For both the Pauli and Dirac equations,
however, each component of $M$, the total
angular momentum operator, commutes with the Hamiltonian $H$, implying that
$M^2$ commutes with $H$ as well.
For this reason, it is conventional to choose
eigenstates of $H$, $M^2$ and $M_z$, with eigenvalues $E_n$,
$\hbar^2 j(j+\half)$ and $\hbar m$, respectively.

There is one further subtlety:  Although the orbital angular momentum
does not commute with the hamiltonian in the Dirac case, it can be
shown that $\ell$ is `almost' a good quantum number (see \cite{BeSa}).
That is to say, eigenstates can be found for which
\be
{\bf L}^2 \psi = \hbar^2 \ell(\ell +1)\psi + \hbar^2 w , 
\ee
where the spinor $w$ is negligible. (Its large components
actually vanish.)
Hence for both the
Dirac and Pauli cases considered in this paper, eigenstates are
presented in terms of quantum numbers $n$, $\ell$, $j$ and $m$
for purposes of comparison, even though
$\ell$ is not strictly a good quantum number in the Dirac case.

Finally,
in the following discussions, the time-dependent phase
factor $e^{-iE_nt/\hbar}$ that accompanies the eigenfunctions
in the solution of the time-dependent Pauli and Dirac equations
will be ignored since it contributes nothing to the associated
currents.

\section{PAULI EIGENSTATES}
In this section we examine the Pauli current \eqref{pcurrent} for
some hydrogen atom eigenstates. These eigenstates, two-component
solutions to the Pauli equation, are given by \cite{BeSa}
\begin{equation}\begin{split}\label{peigen}
\psi_{n,\ell,j=\ell+\frac{1}{2},m} &= \frac{1}{\sqrt{2\ell+1}}R_{n\ell}(r)
\begin{pmatrix} {\scriptstyle
\sqrt{\ell+m+\frac{1}{2}}}~Y_{\ell,m-\frac{1}{2}}(\theta,\phi)\\
-{\scriptstyle
\sqrt{\ell-m+\frac{1}{2}}}~Y_{\ell,m+\frac{1}{2}}(\theta,\phi)\end{pmatrix}  \\
\psi_{n,\ell,j=\ell-\frac{1}{2},m} &= \frac{1}{\sqrt{2\ell+1}}R_{n\ell}(r)
\begin{pmatrix} {\scriptstyle
\sqrt{\ell-m+\frac{1}{2}}}~Y_{\ell,m-\frac{1}{2}}(\theta,\phi)\\
{\scriptstyle
\sqrt{\ell+m+\frac{1}{2}}}~Y_{\ell,m+\frac{1}{2}}(\theta,\phi)\end{pmatrix}
				\end{split}
\end{equation}
Here, the $R_{nl}(r)$ are the standard radial wave functions for the hydrogen
atom and the $Y_{l,m\pm\frac{1}{2}}(\theta,\phi)$ are the spherical
harmonics.\footnote{We use the following convention for the relevant functions,
for consistency with \cite{BeSa}:
\begin{align*}
&Y_{\ell m}(\theta,\phi)=\frac{1}{\sqrt{2\pi}}P_{\ell m}(\cos\theta)e^{im\phi}\\
&P_{\ell m}(x)=\sqrt{\frac{2\ell +1}{2}\frac{(\ell -m)!}{(\ell +m)!}}\frac{1}{2^\ell  \ell !}
(1-x^2)^{m/2}\frac{d^{\ell +m}}{dx^(\ell +m)}(x^2-1)^\ell , ~~~ m\geq 0 \\
& P_{\ell ,-m}(x)=(-1)^m P_{\ell m}(x), ~~~~ m < 0.
\end{align*}}
We use spherical polar coordinates in which $r$ is the radius, $\phi$ is the
angle measured counterclockwise from the $x$-axis and $\theta$ is the
angle measured down from the $z$-axis.

As mentioned earlier, the wave functions given in \eqref{peigen} are
eigenfunctions of $L^2$ (the orbital angular momentum) with
eigenvalue $\hbar^2 \ell(\ell+1)$,
$M^2$ (the total angular momentum) with eigenvalue $\hbar^2 j(j+1)$ and
$M_z$ with eigenvalue $m\hbar$.   In general, however, they are
{\em not} eigenstates of $s_z$, the projection of spin
along the $z$-axis.

The eigenfunctions
can be classified as follows:
\begin{itemize}
\item
For each $n$ value, $\ell$ can assume the values $\ell =0,1,...,n-1$.
\item
For each $\ell $ value, $m$ can assume the values
$m=-\ell +1/2, -\ell +3/2,...,\ell -1/2$.
\item
For each of the above there are two
possibilities, $j=\ell +1/2$ and $j=\ell -1/2$.
\end{itemize}
This accounts for all eigenfunctions listed in \eqref{peigen}.

From \eqref{pcurrent}, the two
contributions to the velocity are given by
\begin{equation}\label{va}
{\bf v}_a=\frac{{\bf j}_a}{\rho}= \frac{\hbar}{2m_ei}\frac{(\psid\nabla\psi -
\psi\nabla\psid)}{\psid\psi} = \frac{\hbar}{m_e}\frac{\text{Im}(\psid \nabla
\psi)}{\psid\psi}
\end{equation}
and
\begin{equation}\label{vb}{\bf v}_b=\frac{{\bf
j}_b}{\rho}=\frac{\hbar}{2m_e}\frac{\nabla \times {\bf s}}{\psid\psi},
\end{equation}
where
\be
{\bf s}=\psid \sigma \psi
\ee
is the `spin vector' and $m_e$ is
the mass of the electron.
In order to determine trajectories for the above hydrogen
eigenstates we must compute the velocities ${\bf v}_a$
and ${\bf v}_b$ for the wave functions given in \eqref{peigen}.

We first examine the velocity ${\bf v}_a$ arising from the
Schr\"{o}dinger current ${\bf j}_a$.
Writing
\be
\psi=\frac{1}{\sqrt{2\ell+1}}R_{n\ell}(r)
\begin{pmatrix} v_1(\theta,\phi) \\v_2(\theta,\phi) \end{pmatrix},
\ee  
the term
$\text{Im}\{\psid\nabla\psi\}$
can be shown to be
\begin{multline}
\text{Im}\{\psid\nabla\psi\}=\frac{1}{2\ell+1}\text{Im}{\big\lbrace} R_{n\ell}R_{n\ell}^\prime
(|v_1|^2+|v_2|^2)\er
+ \frac{1}{r}R_{n\ell}^2(v_1^*\frac{\partial v_1}{\partial\theta}+
v_2^*\frac{\partial v_2}{\partial\theta})\eth  \\
+\frac{1}{r\sin\theta}R_{n\ell}^2(v_1^*\frac{\partial v_1}{\partial\phi}+
v_2^*\frac{\partial v_2}{\partial\phi})\eph{\big \rbrace}.\end{multline}
In the above, the only nonzero term comes from the $\eph$
component:
\be
v_k^*\frac{\partial v_k}{\partial\theta}
=i(m\pm\frac{1}{2})|v_k|^2, ~~~~~ k=1,2,
\ee
so that \eqref{va} yields
\begin{equation}\label{vafinal}
{\bf v}_a =
\frac{\hbar}{m_er\sin\theta}{\Big(}m+\half(\frac{|v_2|^2-|v_1|^2}
{|v_1|^2+|v_2|^2}){\Big)}\eph.\end{equation}

It remains to compute ${\bf j}_b$ and the
corresponding velocity ${\bf v}_b$, with
reference to \eqref{vb}. To do this, we first find the
three components of the spin vector ${\bf s} = \psid\sigma\psi$:
\be
\psid \sigma \psi =
\frac{1}{2\ell+1}R_{n\ell}(r)^2{\Big(}2\text{Re}\{v_1^*v_2\},
2\text{Im}\{v_1^*v_2\}, |v_1|^2-|v_2|^2{\Big)}.
\ee
From the form of the wave functions \eqref{peigen},
\be
2\text{Re}(v_1^*v_2)=2c_1c_2N_1N_2P_\ell^{m-\half}(\theta)P_\ell^{m+\half}(\theta)\cos\phi
\ee
and
\be
2\text{Im}(v_1^*v_2)=2c_1c_2N_1N_2P_\ell^{m-\half}(\theta)P_\ell^{m+\half}(\theta)\sin\phi,
\ee
where the $c_i$ are given in \eqref{peigen}  and
the $N_i$ are the normalization constants of the relevant spherical
harmonics. For simplicity of notation, define
\be
a= c_1N_1P_\ell^{m-\half}, ~~~ b= c_2N_2 P_\ell^{m+\half}. 
\ee
If we write
\be
\label{spinw}
\psid\sigma\psi =
\frac{1}{2\ell+1}R_{n\ell}^2(r){\bf w},
\ee
then the vector ${\bf w}$
can be expressed in Cartesian form as
\begin{equation}\label{wcart} (w_x, w_y, w_z)=(2ab\cos\phi, 2ab\sin\phi,
a^2-b^2). \end{equation} %because a^2 is still |v_1|^2, same for b.
We may also express ${\bf w}$ in spherical polar form, i.e.,
\begin{equation}\label{wpol}
\begin{split}
w_x &= r_s\sin\theta_s\cos\phi_s \\
w_y &= r_s\sin\theta_s\sin\phi_s \\
w_z &= r_s\cos\theta_s,
\end{split}\end{equation}
where the orientation of the spin vector ${\bf s}=\psid\sigma\psi$ is
given by the angles $\theta_s$ and $\phi_s$. 
(See also \cite{BeSa}, p. 62-63 for
a brief discussion of the spin vector.)
A comparison of \eqref{wpol} with \eqref{wcart} suggests
that we might let $r_s=a^2+b^2$, $\phi_s=\phi$
and then compute $\theta_s$ in terms of $\theta$ using the relations
\begin{equation}\label{thes_both}
\cos\theta_s=\frac{a^2-b^2}{a^2+b^2}, ~~~~
\sin\theta_s=\frac{2ab}{a^2+b^2}.
\end{equation}
However, this is consistent with the definition of spherical
coordinates only if $2ab \geq 0$ since $\theta_s$ is restricted to the
interval $[0,\pi]$. When this condition is not met, i.e., $2ab < 0$,
then the polar coordinates for ${\bf w}$ are given by
 $r_s=a^2+b^2$,  $\phi_s=\phi+\pi$ and
\be
\cos\theta_s=\frac{a^2-b^2}{a^2+b^2}, ~~~~
\sin\theta_s=\frac{|2ab|}{a^2+b^2}=-\frac{2ab}{a^2+b^2}. 
\ee

In either of the above cases, the spin vector ${\bf s}$ lies in a plane
defined by the position vector ${\bf r}$ and the $z$ axis, which is in
agreement with \cite{BeSa}.
After some manipulation, we find that
\be
{\bf s} = s_r\er + s_\theta \eth ,
\ee
where
\begin{equation}\label{srsthe}
\begin{split}
s_r &= s\cos\theta(\frac{a^2-b^2}{a^2+b^2})+s\sin\theta
(\frac{2ab}{a^2+b^2}) \\
s_\theta &=
-s\sin\theta(\frac{a^2-b^2}{a^2+b^2})+s\cos\theta(\frac{2ab}{a^2+b^2}).
\end{split}
\end{equation}
(Here $\er$, $\eth$ and
$\eph$ are the spherical polar unit vectors at the position of the
electron.)
Evolution of the position coordinates as the particle follows the
trajectory implies that the spin vector precesses about the $z$-axis.
This was originally described by Holland \cite{Ho93}.

From the above result we find that
\begin{equation}\label{vbfinal}
{\bf v}_b=\frac{\hbar}{2m_e}\frac{\nabla \times {\bf s}}{\psid\psi}
=\frac{\hbar}{2mrs}{\Big(}s_\theta+
r\frac{\partial s_\theta}{\partial r}-
\frac{\partial s_r}{\partial\theta}{\Big)}\eph.\end{equation}
In other words, as was the case for $\va$ in \eqref{vafinal},
the contribution to the velocity from
$\vb$ is again only in the $\eph$ direction.
Therefore, for all eigenstates of the form in \eqref{peigen},
the motion of the electron is in the $\eph$ direction,
i.e. rotational motion about the $z$ axis. This is in qualitative
agreement with the \sch results.

The total speed in the $\eph$ direction is given by
$v = v_a + v_b$, i.e.,
\begin{equation}
v = \frac{\hbar}{m_er\sin\theta}
{\Big(}m+\half(\frac{|v_2|^2-|v_1|^2}{|v_1|^2+|v_2|^2}){\Big)}
+
\frac{\hbar}{2mrs}{\Big(}s_\theta+
r\frac{\partial s_\theta}{\partial r}-
\frac{\partial s_r}{\partial\theta}{\Big)}.
\end{equation}

In what follows it will be useful to understand the relationship
between the velocities for positive and negative (corresponding)
values of $m$.  Recall that for $j=\ell+\half$,
\be
\psi_{n,\ell,j=\ell+\frac{1}{2},m}= \frac{1}{\sqrt{2\ell+1}}R_{n\ell}(r)
\begin{pmatrix} {\scriptstyle
\sqrt{\ell+m+\frac{1}{2}}}~Y_{\ell,m-\frac{1}{2}}(\theta,\phi)\\
-{\scriptstyle
\sqrt{\ell-m+\frac{1}{2}}}~Y_{\ell,m+\frac{1}{2}}(\theta,\phi)\end{pmatrix}
\ee
and 
\be
Y_{\ell m}=\frac{1}{\sqrt{2\pi}}P_{\ell m}e^{im\phi}, ~~~
P_{\ell,-m}(x)=(-1)^m P_{\ell m}(x).
\ee
Also, from the derivation above, the spin vector {\bf s} is
proportional to
\be
{\bf w}=(2\text{Re}\{v_1^*v_2\}, 2\text{Im}\{v_1^*v_2\},
|v_1|^2-|v_2|^2). 
\ee
When $m$ is replaced by $-m$, we have (denoting the new term with a
superscript ${\scriptstyle(-)}$ and the old with ${\scriptstyle(+)}$)
\be
 v_1^{(-)} = {\scriptstyle\sqrt{\ell-m+\half}}Y_{\ell,-m-\half}
= {\scriptstyle\sqrt{\ell-m+\half}}\frac{1}{\sqrt{2\pi}}
(-1)^{m+\half}P_{\ell,m+\half}e^{i(-m-\half)\phi} = v_2^{*(+)}. 
\ee
Similarly,
\begin{align}
 v_2^{(-)} &= -{\scriptstyle\sqrt{\ell+m+\half}}\frac{1}{\sqrt{2\pi}}
P_{\ell,-m+\half}e^{i(-m+\half)\phi} \\
&=  -{\scriptstyle\sqrt{\ell+m+\half}}\frac{1}{\sqrt{2\pi}}(-1)^{m-\half}
P_{\ell,m-\half}e^{-i(m-\half)\phi} =-v_1^{*(+)}. \end{align}
Therefore,
\be
(|v_1|^2-|v_2|^2)^{(-)} = -(|v_1|^2-|v_2|^2)^{(+)}, 
\ee
and
furthermore,
\be
(v_1^*v_2)^{(-)}=v_2^{(+)}(-v_1^{*(+)})= (-v_1^*v_2)^{(+)}. 
\ee
All three components of ${\bf w}$ change sign when $m$ is replaced
with $-m$ (the other eigenvalues are left unchanged), so that the
spin vector in \eqref{spinw} changes sign.  Therefore,
\be
\va^{(-)}=\frac{\hbar}{m_er\sin\theta}{\Big(}m+
\half(\frac{(|v_2|^2-|v_1|^2)^{(-)}}
{(|v_1|^2+|v_2|^2)^{(-)}}{\Big)}\eph = -\va^{(+)}
\ee
and
\be
\vb^{(-)}=
\frac{\hbar}{2m_e}\frac{\nabla \times {\bf s}^{(-)}}{\psid\psi}
= -\vb^{(+)}.
\ee

Thus, both $\va$ and $\vb$ change sign when $m$ changes sign, so that
the overall velocity simply changes direction. This simplifies the
computation of the rates of revolution. A similar proof holds
for the case $j=\ell-\half$.

Before concluding this section, we mention that in their treatment
of the Pauli equation using Euler angles, Bohm and Schiller \cite{BoSch}
(p. 80) deduced that the electron in a
hydrogen atom eigenstate would execute circular
motion about the principal axis with constant angular velocity.
However, no angular velocities were computed in the paper.
In the next section we compute the angular velocities for
the first few Pauli hydrogen eigenstates.

\subsection{ANGULAR VELOCITIES FOR $n=1$ and $n=2$ PAULI EIGENSTATES}
We have computed explicitly
the rates of revolution $d\phi/dt$
for the first few Pauli hydrogen eigenstates following the procedure
described above.
In each case, one computes the velocity $\va$,
followed by the spin vector $s=\psid\psi$,
finding $s_r$ and $s_\theta$ from \eqref{srsthe}.  Then
$\vb$ is computed to give the total velocity {\bf v}.
Since, for all cases, {\bf v} points in the $\eph$ direction,
the angular velocity $d\phi/dt$ is given by
\begin{equation}
\frac{d\phi}{dt}=\frac{v}{r\sin\theta}. \end{equation}
The results of our computations are presented in Table
\ref{pauliresults}.

The first three results presented in Table \ref{pauliresults}
correspond to wave functions that are also eigenstates of
$s_z$ because of the special coupling of spin and orbital
angular momentum vectors.  As expected, these rates of revolution
agree with
those computed in \cite{CoVr} for, respectively, the 
$1s$, $2s$ and $2p_0$ \sch eigenstates.
However, the final two states in Table \ref{pauliresults} are
not spin eigenstates.  As such, they have no
analogue in the \sch case so that no comparisions of rates can be made.

\begin{table}
\begin{tabular}{|>{\hspace{1cm}}c<{\hspace{1cm}}|>{\hspace{1.5cm}}c<{\hspace{1.5cm}}|}
\hline
{\bf Quantum Number} $n$, $\ell$, $j$, $m$ &\rule[-0.4cm]{0cm}{1cm}{\bf Rotation rate} $d\phi/dt$ \\
\hline
 $1,0, \half,\pm\half$ & \rule[-0.5cm]{0cm}{1cm}$\pm\frac{\hbar}{m_ear}$ \\
\hline
$2,0, \half,\pm\half$ & \rule[-0.5cm]{0cm}{1cm}$\pm\frac{\hbar}{2m_ear}
(\frac{1}{1-\frac{r}{2a}}+1)$ \\
\hline
$2,1,\frac{3}{2},\pm\frac{3}{2}$  & \rule[-0.5cm]{0cm}{1cm}$\pm\frac{\hbar}{2m_ear}$
 \\ \hline
$2,1,\half,\pm\half$ & \rule[-0.5cm]{0cm}{1cm}$\pm\frac{\hbar}{m_e r^2}(3-\frac{r}{2a})$ \\
\hline
$2,1,\frac{3}{2},\pm\half$  & \rule[-0.5cm]{0cm}{1cm}$\pm\frac{\hbar}{2m_era}
\frac{8\cos^2\theta-\sin^2\theta}{4\cos^2\theta+\sin^2\theta}$ \\
\hline

\end{tabular}
\caption{Angular rates of revolution for Pauli eigenstates}\label{pauliresults}
\end{table}

\section{DIRAC EIGENSTATES}
We now consider the 4-component Dirac eigenstates for hydrogen.
Following \cite{BeSa}, they are given as follows:
For $j=\ell+\half$,
\begin{equation}\begin{split}\label{estatesp}
\psi_1 &= g(r)\sqrt{\frac{\ell+m+\half}{2\ell +1}}
Y_{\ell,m-\half}(\theta,\phi)  \\
\psi_2 &= -g(r)\sqrt{\frac{\ell-m+\half}{2\ell +1}}
Y_{\ell,m+\half}(\theta,\phi)  \\
\psi_3 &= -i f(r)\sqrt{\frac{\ell-m+\frac{3}{2}}{2\ell +3}}
Y_{\ell+1,m-\half}(\theta,\phi)  \\
\psi_4 &= -i f(r)\sqrt{\frac{\ell+m+\frac{3}{2}}{2\ell +3}}
Y_{\ell+1,m+\half}(\theta,\phi)
				\end{split}
\end{equation}
and for $j=\ell-\half$,
\begin{equation}\begin{split}\label{estatesm}
\psi_1 &= g(r)\sqrt{\frac{\ell-m+\half}{2\ell +1}}
Y_{\ell,m-\half}(\theta,\phi)  \\
\psi_2 &= g(r)\sqrt{\frac{\ell+m+\half}{2\ell +1}}
Y_{\ell,m+\half}(\theta,\phi)  \\
\psi_3 &= -i f(r)\sqrt{\frac{\ell+m-\half}{2\ell -1}}
Y_{\ell-1,m-\half}(\theta,\phi)  \\
\psi_4 &= i f(r)\sqrt{\frac{\ell-m -\half}{2\ell -1}}
Y_{\ell-1,m+\half}(\theta,\phi).
				\end{split}
\end{equation}
The $Y_{\ell,m}$ are the usual spherical harmonics
and $f(r)$ and $g(r)$ are the normalized radial Dirac eigenfunctions
(see \cite{BeSa} p. 69). Recall that
even though $\ell$ is not a good quantum
number, the eigenstates are written in terms of $\ell$ because it is
`almost' a good quantum number, and also because these solutions to
the Dirac equation are built from the corresponding Pauli
eigenstates. (For a complete discussion, see \cite{BeSa}.)

\subsection{TRAJECTORIES FOR GENERIC DIRAC EIGENSTATES}
In this section, we show that Bohm trajectories for Dirac hydrogen
share common features.
First, the components of the Dirac current in \eqref{dcurrent}
may be expressed in terms of the
components of the wave function as follows,
\begin{equation}\begin{split}\label{currents}
\frac{1}{c}j_x &=
2\text{Re}\{\psid_1\psi_4\}+2\text{Re}\{\psid_2\psi_3\}\\
\frac{1}{c}j_y &=
2\text{Im}\{\psid_1\psi_4\}-2\text{Im}\{\psid_2\psi_3\}\\
\frac{1}{c}j_z &=
2\text{Re}\{\psid_1\psi_3\}-2\text{Re}\{\psid_2\psi_4\}.
				\end{split}
\end{equation}
We now compute these components using the hydrogenic wave
functions given in \eqref{estatesp} and \eqref{estatesm}.  Starting
with the $z$ component, we find that in the $j=\ell+\half$ case,
\be
\psid_1\psi_3=-i f(r)g(r)\sqrt{\frac{\ell+m+\half}{2\ell +1}}
\sqrt{\frac{\ell-m+\frac{3}{2}}{2\ell +3}}
Y_{\ell,m-\half}(\theta,\phi)^*
Y_{\ell+1,m-\half}(\theta,\phi) 
\ee
and
\be
\psid_2\psi_4= if(r)g(r)\sqrt{\frac{\ell-m+\half}{2\ell +1}}
\sqrt{\frac{\ell+m+\frac{3}{2}}{2\ell +3}}
Y_{\ell,m+\half}(\theta,\phi)^*
Y_{\ell+1,m+\half}(\theta,\phi). 
\ee
Both $ \psid_1\psi_3$ and $\psid_2\psi_4$ are imaginary since the phases
of the spherical harmonics cancel, implying that $j_z=0$.
This is also the case for
$j=\ell-\half$.
Therefore, in all cases,
motion of the electron is constrained to planes of constant
$z$.  While this is a simple result,  it applies
to all hydrogen eigenstates of the forms \eqref{estatesp} and
\eqref{estatesm} and is therefore of general interest.

We find the other components of the current in a similar fashion.
For the $j=\ell+\half$ case,
\be
\label{jx}
\frac{1}{c}j_x= 2\sin\phi f(r)g(r)\left(
{\scriptstyle \sqrt{\frac{\ell-m+\half}{2\ell +1}}
\sqrt{\frac{\ell-m -\half}{2\ell -1}}}
P_{\ell,m-\half}P_{\ell+1,m+\half}
+{\scriptstyle \sqrt{\frac{\ell-m+\half}{2\ell +1}}
\sqrt{\frac{\ell-m+\frac{3}{2}}{2\ell +3}}}
P_{\ell,m+\half}  P_{\ell+1,m-\half} \right).
\ee
We define $F(\cos\theta)$ to be the quantity in brackets so that
\be
\frac{1}{c}j_x = 2\sin\phi f(r)g(r)F(\cos\theta).
\ee
Because of its similarity in form, $j_y$ also has
the $F(\cos\theta)$ term:
\be
\frac{1}{c}j_y=-2\cos\phi f(r)g(r)F(\cos\theta).
\ee
From \eqref{currvel} and \eqref{dcurrent},
the motion of the electron in a plane of constant $z$ is given
by the following system of DEs:
\begin{equation}\begin{split}\label{des}
\dot{x} &= \frac{j_x}{\psid\psi}=
\frac{2cf(r)g(r)F(\cos\theta)\sin\phi}{\psid\psi} \\
\dot{y} &= \frac{j_y}{\psid\psi}=
\frac{-2cf(r)g(r)F(\cos\theta)\cos\phi}{\psid\psi}.
				\end{split}
\end{equation}
From the polar forms of $x$ and $y$, we have 
\be
x\dot{x}+y\dot{y}= \frac{d}{dt}(x^2+y^2) = 0 .
\ee
In other words, the motion is circular about
the $z$-axis.

A similar proof applies to the $j=\ell-\half$ case. For the sake of the
computations in the next section, the components of the current
for this case are $j_z=0$,
\begin{equation}\begin{split}\label{jx2}
\frac{1}{c}j_x &= -2\sin\phi f(r)g(r)\left(
{\scriptstyle
\sqrt{\frac{\ell-m+\half}{2\ell +1}}\sqrt{\frac{\ell-m -\half}{2\ell
-1}}}
P_{\ell,m-\half} P_{\ell-1,m+\half}
+{\scriptstyle\sqrt{\frac{\ell+m+\half}{2\ell +1}}
\sqrt{\frac{\ell+m-\half}{2\ell
-1}}}P_{\ell,m+\half}P_{\ell-1,m-\half}
\right) \\
&=-2\sin\phi f(r)g(r) G(\cos\theta) \end{split}\end{equation}
 and
\begin{equation}
\frac{1}{c}j_y = 2\cos\phi f(r)g(r)G(\cos\theta),\end{equation} where
$G(\cos\theta)$ is the term in brackets in
\eqref{jx2}.  Once again we find that motion about the $z$-axis is circular.

In summary, we have shown that electron trajectories associated with
Dirac hydrogen eigenstates are circular, as was the case
for Pauli eigenstates.
In the next section, we compute some rates of revolution
for these trajectories and their nonrelativistic limits.

Finally, note that if $m$ changes from positive to negative, both
$F(\cos\theta)$ and $G(\cos\theta)$ simply undergo an overall sign
change. This means that the angular rotation simply changes direction,
but maintains the same functional form for $m=\pm\half$, as was
the case for the Pauli trajectories.

\subsection{ANGULAR VELOCITIES FOR $n=1$ and $n=2$ DIRAC EIGENSTATES}
We find the angular rate of revolution in general from \eqref{des}
using the relation
\be
\dot{y}=(r\sin\theta\cos\phi)\dot{\phi}. 
\ee
From \eqref{des}, it follows that
\begin{equation}\label{dphidt}
\frac{d\phi}{dt}= -\frac{2cf(r)g(r)F\cos(\theta)}{\psid\psi
r\sin\theta\cos\phi}. \end{equation}
Although this equation is deceptively simple in appearance,
the functions $f(r)$, $g(r)$ -- and therefore
$\psid\psi$ -- are quite complicated in form.  Since we already know the
qualititative motion, explicit computations of the rates of revolution
for the general case, beyond the result given in \eqref{dphidt}, are
not particularly enlightening.

However, it is
enlightening is to examine the nonrelativistic limits of
\eqref{dphidt} and compare the results to the values computed from the
Pauli equation. If the de Broglie-Bohm picture is to give a coherent account of
the hydrogen atom, these results must agree.  In what follows, we examine the
nonrelativistic limits of \eqref{dphidt} for the $n=1$ and $n=2$
eigenstates.  Note that we compute only the positive $m$ value since
the angular velocity simply changes sign for negative $m$.

\subsubsection{n=1}
In the ground state, we have \cite{BeSa}
\begin{equation}\begin{split}
g(r) &= \left(\frac{2}{a}\right)^{3/2}
\sqrt{\frac{1+\epsilon_1}{2\Gamma(2\gamma_1+1)}}
e^{-\rho_1/2}\rho_1^{\gamma_1-1} \\
f(r) &= -\sqrt{\frac{1-\epsilon_1}{1+\epsilon_1}}g = -\delta g ,
				\end{split}
\end{equation}
where
\begin{equation}\label{defns1}
 \gamma_1=\sqrt{1-\alpha^2},~~~~ \rho_1=2r/a,
~~~~\epsilon_1=\left(1+\frac{\alpha^2}{\gamma_1^2}\right)^{-1/2},
~~~~\delta=\sqrt{\frac{1-\epsilon_1}{1+\epsilon_1}}, \end{equation}
 $\alpha$ is the fine structure constant, and $a$ is the Bohr radius.
The ground state wave function is given by
\be
\psi_1=\frac{g}{\sqrt{4\pi}}, ~~~\psi_2=0,
~~~\psi_3=-\frac{1}{\sqrt{4\pi}}if \cos\theta,
~~~\psi_4=-\frac{1}{\sqrt{4\pi}}if\sin\theta e^{i\phi}. 
\ee
This gives
\be
\rho = \psid\psi = \frac{1}{4\pi}(1+\delta^2)g^2~~~ \text{and}~~~
F(\cos\theta)=\frac{1}{4\pi}\sin\theta.
\ee
Substitution into \eqref{dphidt} yields, after cancellation,
\begin{equation}\label{phi1}
 \frac{d\phi}{dt}={\big(}\frac{2}{r}{\big)}\frac{\delta
 c}{1+\delta^2}.
\end{equation}
In the nonrelativistic limit, $c \rightarrow \infty$,
which implies that $\alpha = e^2/\hbar c \rightarrow 0$ and
$\gamma_1\rightarrow 1$.
Furthermore, this implies that $\epsilon_1\rightarrow 1$ and
$\delta \rightarrow 0$.
In order to determine the behaviour of $\delta c$, we expand $\epsilon_1$ as
\be
\epsilon_1 \approx 1-\half(\frac{\alpha^2}{\gamma_1^2})
\ee
so that
\be
\frac{1-\epsilon_1}{1+\epsilon_1}\rightarrow \frac{1}{4}
\alpha^2~~~\text{as}~~~c\rightarrow\infty .
\ee
Therefore $\delta\rightarrow \half\alpha$.
Substitution into \eqref{phi1} yields
\be
\frac{d\phi}{dt}\rightarrow\frac{1}{r}\alpha c = \frac{e^2}{r\hbar},
\ee
which, when written in terms of the Bohr radius $a$, becomes
\begin{equation}\label{1srate}\frac{d\phi}{dt}=
\frac{\hbar}{m_ear}.\end{equation}
This is the angular rotation rate for the ground state Dirac wave
function. It is in agreement with the \sch rate given by Holland
\cite{Ho93}, and it is also in agreement with the rate found for
the Pauli equation in Table \ref{pauliresults}.

\subsubsection{n=2}
\begin{enumerate}
\item{$2S_{1/2}$ ($n=2,~~\ell=0,~~j=\frac{1}{2},~~ m=\frac{1}{2}$)}

As in the $1s$ case, we have the wave function
\be
\psi_1=\frac{g}{\sqrt{4\pi}}, ~~~\psi_2=0,
~~~\psi_3=-\frac{1}{\sqrt{4\pi}}if \cos\theta,
~~~\psi_4=-\frac{1}{\sqrt{4\pi}}if\sin\theta e^{i\phi},
\ee
where the functions $f(r)$ and $g(r)$ are suitably modified for the
$n=2$ case. Again, their exact functional form is not relevant, but
the relationship between $f$ and $g$ is important; here, rather than
\eqref{defns1} we have
\begin{equation}\label{defns2}
\rho_2=\frac{2r}{N_2 a}, ~~~ N_2=\sqrt{2(1+\gamma_1)},~~~
\epsilon_2=\left(1+(\frac{\alpha}{1+\gamma_1})^2\right)^{-1/2}, ~~~
\delta=\sqrt{\frac{1-\epsilon_2}{1+\epsilon_2}}A, \end{equation}
and the number $A$ is given by
\be
A=\frac{(2\gamma_1+1)(N_2+2)-(N_2+1)\rho_2}{(2\gamma_1+1)N_2
-(N_2+1)\rho_2}. 
\ee
After some cancellation, substitution of the wave function into
\eqref{dphidt} gives, as in \eqref{phi1},
\begin{equation}\label{phi2}
 \frac{d\phi}{dt}={\big(}\frac{2}{r}{\big)}\frac{\delta
 c}{1+\delta^2}.
\end{equation}
Once again, we examine how the quantities in \eqref{defns2} behave in
the nonrelativistic limit $c\rightarrow\infty$.  In this case,
\be
\epsilon_2 \approx 1-\half (\frac{\alpha}{1+\gamma_1})^2 
\ee
so that
\be
\sqrt{\frac{1-\epsilon_2}{1+\epsilon_2}}\rightarrow
\frac{\alpha}{4}.
\ee

From the properties $\gamma_1\rightarrow 1$, hence  $N_2\rightarrow2$,
the limit of $A$ in \eqref{defns2} is
\be
A \rightarrow \frac{4-\rho_2}{2-\rho_2}. 
\ee
This implies that
\be
\delta c \rightarrow \alpha c \frac{4-\rho_2}{4(2-\rho_2)}
\ee
and \eqref{phi2} becomes
\be
\frac{d\phi}{dt}= \frac{ e^2 (4-\rho_2)}{2\hbar r(2-\rho_2)},
\ee
which can be rewritten as
\begin{equation}
\frac{d\phi}{dt}=\frac{\hbar}{2m_ear}\left(1-\frac{1}{\frac{r}{2a}-1}\right).
\end{equation}
This is the angular rotation rate for the $2s$ \sch state given
in \cite{CoVr} and is also in agreement with the Pauli result of
Table 1.
\item{$2P_{3/2}$ ($n=2,~~\ell=1,~~j=\frac{3}{2},~~ m=\frac{3}{2}$)}

In this case the wave function is given by
\begin{equation}
\psi_1=\sqrt{\frac{3}{8\pi}}g\sin\theta e^{i\phi}, ~~~\psi_2=0,~~~
\psi_3=-if\sqrt{\frac{3}{8\pi}}\cos\theta\sin\theta e^{i\phi},
~~~\psi_4=-if\sqrt{\frac{3}{8\pi}}\sin^2\theta e^{3i\phi}.\end{equation}
The functions $f$ and $g$, and the relationship between them, will be
the same as in the above case, because only $m$ has changed. However,
we now have $\psid\psi= \frac{3}{8\pi}\sin^2\theta g^2$ in the limit
as $\delta \rightarrow 0$. Therefore, the expression corresponding to
\eqref{phi1} is
\begin{equation}
 \frac{d\phi}{dt}= \frac{-2c\delta}{r}. \end{equation}
Substitution of the nonrelativistic limiting expressions from the
previous case yields
 \begin{equation}
 \frac{d\phi}{dt}= \frac{\hbar}{2m_ear}.\end{equation}
This is the angular rotation rate for the $2p_1$ \sch state
given in \cite{CoVr} and is also in agreement with the Pauli
result in Table 1.

\item{$2P_{1/2}$ ($n=2,~~\ell=1,~~j=\frac{1}{2},~~ m=\half$)}

This case is similar to the previous one although the functional forms
of $f$ and $g$ are different. We have
\begin{equation}
\psi_1=\frac{1}{\sqrt{4\pi}}g\cos\theta,
~~~\psi_2=\frac{1}{\sqrt{4\pi}}g\sin\theta e^{i\phi},~~~
\psi_3=-if\frac{1}{\sqrt{4\pi}},
~~~\psi_4=0\end{equation}
and
\be
\frac{d\phi}{dt}={\big(}\frac{2}{r}{\big)}\frac{\delta
 c}{1+\delta^2} 
\ee
and most of the definitions of \eqref{defns2}
 remain the same.  In this case, the term $A$ is given by
\be
A=\frac{(2\gamma_1+1)N_2-(N_2-1)\rho_2}{(2\gamma_1+1)(N_2-2)
-(N_2-1)\rho_2}.
\ee
In the non-relativistic limit,
\be
A \rightarrow \frac{6-\rho_2}{-\rho_2} 
\ee 
so that
\be
\frac{d\phi}{dt}=-\frac{\hbar}{m_er^2}\left(3-\frac{r}{2a}\right).
\ee
In this case, no comparison can be made with any \sch state.
\item{$2P_{3/2}$ ($n=2,~~\ell=1,~~j=\frac{3}{2},~~ m=\half$)}

This case is somewhat different.  Here, the wave function is
\begin{equation}\begin{split}
\psi_1=\frac{1}{\sqrt{2\pi}}g\cos\theta,
~~~&\psi_2=-\frac{1}{\sqrt{8\pi}}g\sin\theta e^{i\phi} \\
\psi_3=-if\frac{1}{\sqrt{8\pi}}(3\cos^2\theta-1),
~~~&\psi_4=-if\sqrt{\frac{9}{8\pi}}\sin\theta\cos\theta e^{i\phi}
				\end{split}
\end{equation}
and
\be
\psid\psi = \frac{1}{8\pi}(4\cos^2\theta+\sin^2\theta)g^2
+\frac{1}{8\pi}((3\cos^2\theta-1)^2+3\sin^2\theta\cos^2\theta)f^2. 
\ee
Furthermore, the function $F(cos\theta)$ becomes
\be
F(cos\theta) = \frac{1}{8\pi}\sin\theta(8\cos^2\theta-\sin^2\theta).
\ee
Again, $f=-\delta g$, with
\begin{equation}\label{defns3}
\delta=\sqrt{\frac{1-\epsilon_3}{1+\epsilon_3}},~~~~
\epsilon_3=\left(1+\frac{\alpha^2}{\gamma_2^2}\right)^{-1/2},~~~~
\gamma_2=\sqrt{4-\alpha^2}\end{equation}
and we find from \eqref{dphidt} that
\begin{equation}\label{philim}
 \frac{d\phi}{dt}= \frac{2\delta
c(8\cos^2\theta-\sin^2\theta)}{r(4\cos^2\theta+\sin^2\theta)}.\end{equation}
As $c\rightarrow \infty$,
\be
\epsilon_3\approx 1-\half(\frac{\alpha}{2})^2 
\ee
so that
$\delta c\rightarrow \frac{\alpha}{4}$.
After substituting and rewriting $\alpha$, we obtain
\begin{equation}
 \frac{d\phi}{dt}=\frac{\hbar}{2m_ear}
{\big(}\frac{8\cos^2\theta-\sin^2\theta}
{4\cos^2\theta+\sin^2\theta}{\big)}.\end{equation}
Once again, no comparison can be made with any \sch state.
\end{enumerate}

In each case presented above, the nonrelativistic limit of the Dirac
angular velocity agrees with the corresponding Pauli result given in Table
\ref{pauliresults}. We expect this, since the Pauli equation is the
nonrelativistic limit of the Dirac equation.   However, the
results are not obvious,
since the expressions in  
\eqref{dcurrent} and \eqref{pcurrent} for, respectively,
the Dirac and Pauli currents are
quite different.

\section{CONCLUDING REMARKS}
In this paper, we have
determined the general features of de Broglie-Bohm trajectories
for energy/total angular momentum eigenstates 
of the Pauli and Dirac hamiltonians for hydrogen.
In all cases, the electron, assumed to be in an eigenstate of
$M_z$, the $z$-component of the total angular momentum, $M$,
is confined to a plane of constant $z$-value and
executes circular motion about the $z$-axis 
with a constant angular velocity $d\phi/dt$.  
As well, we have outlined a procedure to
compute these angular velocities for general
eigenstates and have explicitly computed them for
the $n=1$ and $n=2$ Pauli and Dirac
hydrogen eigenstates.  

In the cases where the Pauli eigenstates are also eigenstates of the
$s_z$ operator, our results from the Pauli equation agree with earlier
computations of the trajectories of corresponding Schr\"{o}dinger eigenstates
\cite{CoVr}. Furthermore, the nonrelativistic limits of the Dirac
results agree with the Pauli results.  We have therefore shown that the de
Broglie-Bohm causal picture can be applied coherently to the hydrogen
atom, moving from the \sch to the Pauli and ultimately to the Dirac equation.

Finally, one may well wish to consider trajectories for 
Pauli or Dirac wave functions other than those considered in this
paper.  For example, it may be interesting to examine trajectories
for particular linear combinations of eigenstates.  (In \cite{CoVr},
we examined Bohm trajectories for the familiar \sch $2p_x$ and $2p_y$ 
orbitals used in descriptions of chemical bonding.
As well, we examined
trajectories associated with a time-varying linear combination 
of $1s$ and $2p_0$ hydrogenic wave functions that simulated an electronic
transition induced by an oscillating electric field.)
The method of computing Bohm trajectories outlined in 
Sections 2 and 3 can be extended in a straightforward manner
to treat such linear combinations, although the computations
may well become quite complicated.

\section*{ACKNOWLEDGEMENTS} We gratefully acknowledge that this
research has been supported by the Natural Sciences and Engineering
Research Council of Canada (NSERC) in the form of a Postgraduate
Scholarship (CC) and an Grant in Aid of Research
(ERV). CC also acknowledges partial financial support from the
Province of Ontario (Graduate Scholarship) as well as the
Faculty of Mathematics, University of Waterloo.

\end{document}